# The General Model of the Cloud-based Learning and Research Environment of Educational Personnel Training

Mariya Shyshkina[1]

[1]Institute of Information Technologies and Learning Tools of NAES of Ukraine,
9 M.Berlynskoho St., Kyiv, Ukraine

{shyshkina}@iitlt.gov.ua

**Abstract.** The article highlights the promising ways of providing access to the cloud-based learning and research software in higher educational institutions. It is emphasized that the cloud computing services implementation is the actual trend of modern ICT pedagogical systems development. The analysis and evaluation of existing experience and educational research of different types of software packages use are proposed. The general model of formation and development of the cloud-based learning and research environment of educational personnel training is substantiated. The reasonable ways of methods selection on the basis of the proposed model are considered and the prospects for their use in educational systems of higher education are described. The analysis and assessment of the prospects of the cloud-based educational and research environment development is fulfilled.

## 1 Introduction

### 1.1 The Problem Statement

Nowadays, innovative technological solutions for learning environment organization and design using cloud computing (CC) and ICT outsourcing have shown promise and usefulness [1-4]. So, the modelling and analysis of the processes of the cloud-based learning environment formation, elaboration and deployment in view of the current tendencies of ICT advance have come to the fore. Insufficient attention to these issues may have the negative impact on the level of ICT competence of lecturers and academic staff, organization of their educational and research activities. There is a need to produce some general framework of learning environment formation and design that may be detailed through the system of the models of its parts so as to elaborate and evaluate different CC based learning components, to consider and compare the most advisable methods of their design and delivery.

## 1.2 The State of the Art

According to the recent research [2], [4]; [6], [9], [14] the problems of cloud technologies implementing in higher educational institutions to provide software access, support collaborative learning, research and educational activities, exchange experience and also project development are especially challenging. The formation of the cloud-based learning environment is recognized as a priority by the international educational community [7], and is now being intensively developed in different areas of education [6], [10], [14].

According to IDC, at the end of 2016, two-thirds of IT executive directors of the 500 largest companies (rated Financial Times) reported that they considered digital transformation as the main objective of corporate strategy. Under the "digital transformation" the gradual transfer of IT infrastructure to the corporate cloud and then partly or completely to the external provider cloud to create more efficient business, accelerate the process of innovation, to achieve cost reduction, improve production safety, etc. is implied [5], [13].

Growing demand for cloud services is reflected by the following figures: the share of those companies in the world that are using the cloud solutions increased from 26% in 2015 to 59% in 2016, representing 127% of annual growth; while in Ukraine this figure of annual growth was 26% (an increase from 38% in 2015 to 48% in 2016) - according to the survey IDC, that was conducted in 2016 with the number of respondents n = 11 350 from all over the world [5], [13].

Among the current issues there are those concerning existing approaches and models for electronic educational resources delivery within the cloud-based setting; the methodology of CC-based learning and research university environment design; evaluation of current experience of cloud-based models and components use [2], [6], [8], [9], [14]. This brings the problem of educational personnel training to the forefront.

## 1.3 The Purpose of the Article

The *main purpose* of the article is to define the general model of the cloud-based learning and research university environment for educational personnel training and consider the possible ways and techniques of its use and application within the pedagogical systems of higher education. The main idea is in the hypothesis that design and development of learning and research environment due to the proposed approach will result in positive effect of better access to electronic educational resources and more efficient use of ICT and also rising of the ICT competence of educational personnel.

## 1.4 The Research Methods

The research method involved analysing the current research (including the domestic and foreign experience of the application of cloud-based learning services to reveal the concept of the investigation and research indicators), examining existing models and approaches, technological solutions and psychological and pedagogical assumptions about better ways of introducing innovative technology so as to consider and elaborate the general model and special methodical system of educational personnel training with the use of the cloud-based components on the basis of Microsoft Office 365, AWS, SageMathCloud. To measure the efficiency of the proposed approach the pedagogical experiment was undertaken. The special indicators to reveal ICT competence of educational personnel trained according to the proposed methodical system within the cloud-based learning environment were used.

## 2  Presenting the Main Material

The use of ICT affects the content, methods and organizational forms of learning and managing educational and research activities that require new approaches to learning environment arrangement [1], [2]. Therefore, the formation of modern cloud-based systems for supporting learning and research activities should be based on appropriate innovative models and methodology that can ensure a harmonious combination and embedding of various networking tools into the educational environment of higher education institution [1], [2], [7], [9].

Taking into account the stages and components of the learning environment design process the general model of this environment formation and development is to be considered (Fig.1).

The *target* component of this model reveals the goals and objectives of the *CC-based learning and research environment* design such as ICT-competent specialist formation, wider access to ICT; providing the use of the most advanced tools and technologies in education and research and others.

The learning and research process is realized within some *Pedagogical System* (PS) being the component of this model. The CC-based environment is designed so as to support the specific aims achievement within this system. For this purpose the certain *functions of PS* should be revealed, among them there are such as learning, research, development, education, control. To fulfill these functions the necessary educational preconditions are to be provided with the use of the CC-based learning environment tools and services.

For the CC-based Learning Environment (LE) would provide the PS functions implementation the specific services of LE are arranged to support the *functions of LE* such as collection, accumulation, storage, input, representation, manipulation and reorganization of data, learning tools management, measurement, communication, maintaining domain Electronic Educational Resources (EER) and others. To deploy the relevant environment components the special *methodical system* is elaborated basing on certain methodological principles, methods and approaches such as open education principles and also the principles being specific only for the cloud-based learning environment in particular adaptability; services personalization; infrastructure unification; full-scale interactivity; flexibility and scalability; consolidation of data and resources and others; the problem-oriented, the person-oriented, operational and processing, holistic and other conceptual approaches are used.

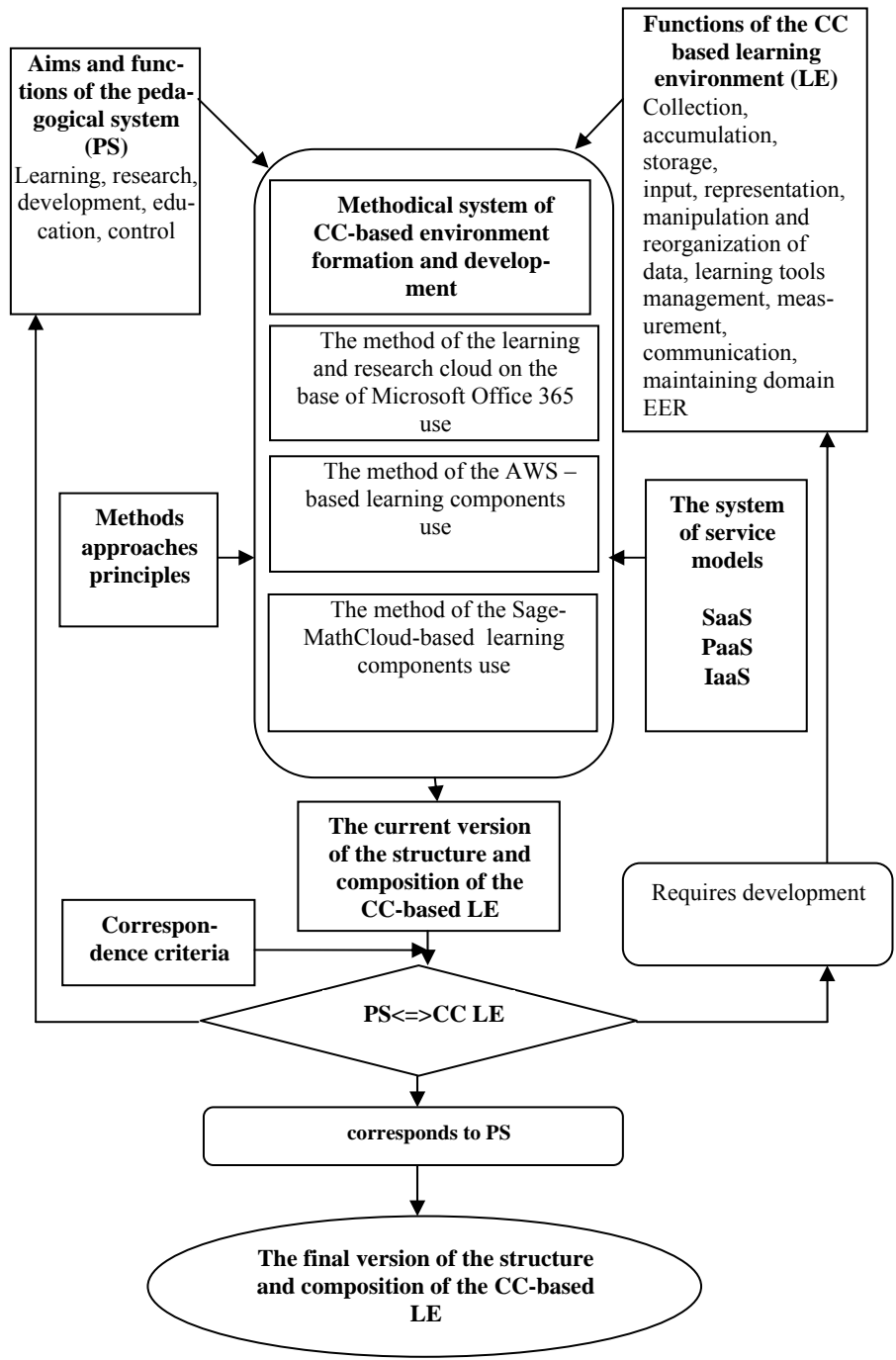

**Fig. 1.** The general model of formation and development of the CC-based learning and research environment of educational personnel training.

The methodical system includes the number of special developed methods that may be realized at 3 levels due to the main types of CC service models such as SaaS (software-as-a-service); PaaS (platform-as-a-service); IaaS (infrastructure-as-a-service) [9]. Conceptual principles, and also the basic characteristics and service models of CC appear to be the system creative factor that brings together the separate methods into the unite system. Due to this the method of the learning and research cloud on the base of Microsoft Office 365 use, the method of the learning components on the basis of AWS use [9], [10], the method of the learning components on the base of SageMathCloud use [6] are elaborated as the methodical system components.

On the basis of the methodical system of the CC-based LE formation and development the current version of its structure and composition is built then this is checked up as for the correspondence to the most complete realization of the PS functions due to the determined criteria. In case of compliance achievement the environment is successfully created. If the structure and composition of LE don't correspond to the functions of PS the environment development is required. For this purpose the return to the review of the environment structure and composition is needed.

**The peculiarities of methods selection for the formation and development of the cloud-based learning environment.** The cloud-based learning environment design basing of the general model of its formation and development involves the methodical system that covers several methods of the learning and research components use that are united by the system creating factor such as the cloud-based approach with the most important basic features and service models. The cloud technologies are to provide the key features of the learning environment that are openness and flexibility. If the tasks of environment development are changed it is possible to modify adequately the tools features and overall composition of the environment and also to modernize the methods of its use. So the structure and the composition can be aligned with the planned development goals, new challenges appeared or appearing in future.

So the methods of the learning and research components use may be selected and elaborated depending on the service model chosen for the environment deployment. For example using the SaaS model the services of the general purpose and also the specialized services may be provided. It may be used to support the learning or research activity regardless of the domain by means of documents, electronic tables, databases, files repositories, sites elaborating and processing and also by means of communication services [4], [7]. On the basis of the general use services the method of the Microsoft Office 365 learning and research cloud use was elaborated.

The specialized services may cover the remote learning laboratories and software tools for programming, design, modeling, data processing, research and others. The clear example of such kind of tools is the SageMathCloud mathematical software [6]. The method of the SageMathCloud learning components use was elaborated within this study. In this case the specialized software is provided as the turnkey solution it does not require deployment on the equipment of a user but still can not be modified in realization proposed by the provider [6].

As opposed to it the IaaS or PaaS model is to provide the software access of whatever kind deployed on the cloud servers of the provider by the user. The method of the learning components use within this kind of model was elaborated on the base of AWS (Amazon Web Services) [9], [10]. This approach may be used to provide access to different kinds of learning services deployed in the cloud. For example the learning component with the Maxima system use was elaborated and researched [10].

**The Results of Experimental Work.** The aim of pedagogical experiment was to justify empirically the proposed theoretical assumptions and methodical system to be really effective.

At the ascertaining stage of the experiment (2012) some problems of ICT competencies of lecturers and academic staff taking part in the experiment as for using CC technologies in learning and research were found. The proportion of those having the high levels of the relevant ICT competencies was 14,65% in the experimental group (58 members) and 12,65% in the control group (60 members). The participants would like to improve their knowledge through greater use of cloud-based tools in their professional activities.

On the formative stage (2012-2014) the different components of the methodical system were introduced in the process of learning and research activities of the experimental group. Comparing the levels of the relevant ICT competencies in the early formative stage and at the end of the experiment gives the rise of the proportion of research and lecturers staff with the high levels of relevant ICT competencies (29,5% in the control group and 48,05% in the experimental group). It shows the statistically relevant increase proved by the angular Fisher criteria ($\varphi_{exp}=2,04>\varphi_{0,05}=1,64$).

Analysis of the formative stage of pedagogical experiment showed that the distribution of the levels of ICT competences in the experimental and the control groups of teaching staff have statistically significant differences due to the introduction of the developed technique, which confirms the hypothesis of the study.

The advantage of the approach is the possibility to compare the different ways to implement resources with regard to the learning infrastructure. Future research in this area should consider different types of resources and environments.

**Conclusions**

The general model of the university cloud-based learning and research environment formation and design proved to be a reasonable framework to deliver and research the cloud-based learning resources and components. The ways of methods selection on the basis of the proposed model and the prospects for their use within the learning systems of higher education are considered. Among them there are the methods of cloud-based components design on the basis of Microsoft Office 365, AWS, SageMathCloud to support learning and research processes. The recent tendencies of CC development are considered and evaluated in view of the cloud-based learning and research environment creation.

## References

1. Bykov, V.: ICT Outsourcing and New Functions of ICT Departments of Educational and Scientific Institutions. In: Information Technologies and Learning Tools, 30(4), http://journal.iitta.gov.ua/index.php/itlt/article/view/717 (2012) (in Ukrainian)
2. Bykov, V., Shyshkina, M.: Emerging Technologies for Personnel Training for IT Industry in Ukraine. In: Interactive Collaborative Learning (ICL), 2014 International Conference on, IEEE, pp. 945-949 (2014)
3. Doelitzscher, F., Sulistio, A., Reich, Ch., Kuijs, H., Wolf, D.: Private cloud for collaboration and e-Learning services: from IaaS to SaaS. In: Computing, 91, pp.23–42 (2011)
4. Lakshminarayanan, R., Kumar, B., Raju, M.: Cloud Computing Benefits for Educational Institutions. In: Second International Conference of the Omani Society for Educational Technology, Muscat, Oman: Cornell University Library, http://arxiv.org/ftp/arxiv/papers/1305/1305.2616.pdf (2013)
5. Middleton, S., G.: Desktop PC Residual Value Forecast. Industry Development and Models. In: IDC US41663116, August 2016, 15 p. (2016)
6. Popel, M., Shokalyuk, S., Shyshkina, M.: The Learning Technique of the SageMathCloud Use for Students Collaboration Support. In: ICT in Education, Research and Industrial Applications: Integration, Garmonization and Knowledge Transfer, CEUR-WS.org/Vol–1844, pp.327-339, (2017)
7. Shyshkina, M. P., Popel, M. V.: The cloud-based learning environment of educational institutions: the current state and research prospects. In: J. Information Technologies and Learning Tools, vol. 37, № 5, http://journal.iitta.gov.ua/index.php/itlt/article/view/1087/829 (2014) (in Ukrainian)
8. Shyshkina, M.: Emerging Technologies for Training of ICT-Skilled Educational Personnel In: Ermolayev, V., Mayr, H.C., Nikitchenko, M., Spivakovsky, A., Zholtkevych, G. (Eds.) ICT in Education, Research and Industrial Applications. Communications in Computer and Information Science, vol. 412, pp. 274–284. Springer, Berlin-Heidelberg (2013)
9. Shyshkina, M.: The Hybrid Service Model of Electronic Resources Access in the Cloud-Based Learning Environment. In: CEUR Workshop Proceedings, vol.1356, pp.295-310, (2015).
10. Shyshkina, M.P., Kohut, U.P., Bezverbnyy, I.A.: Formation of Professional Competence of Computer Science Bachelors in the Cloud Based Environment of the Pedagogical University. In: J. Problems of Modern Teacher Training, vol. 9, part 2, pp.136–146 (2014) (in Ukrainian)
11. Shyshkina, M.P., Spirin, O.M., Zaporozhchenko, Yu.G.: Problems of Informatization of Education in Ukraine in the Context of Development of Research of ICT-Based Tools Quality Estimation. In: J. Information Technologies and Learning Tools, vol. 27, № 1, http://journal.iitta.gov.ua/index.php/itlt/article/view/ 632/483 (2012) (in Ukrainian)
12. Smith, A., Bhogal J., Mak Sharma: Cloud computing: adoption considerations for business and education. In: 2014 International Conference on Future Internet of Things and Cloud (FiCloud), (2014)
13. Turner, M., J.: CloudView Survey 2016: U.S. Cloud Users Invest in Mature Cloud Systems Management Processes and Tools. In: IDC Survey Spotlight US40977916, Jan 2016, (2016)
14. Vaquero, L. M.: EduCloud: PaaS versus IaaS cloud usage for an advanced computer science course. In: IEEE Transactions on Education, 54(4), pp.590–598, (2011)